\documentstyle[psfig,prb,twocolumn,aps,floats]{revtex}

\begin{document}

\draft
\tightenlines
\widetext

\title{Ab-Initio Calculation of the Metal-Insulator Transition in Lithium rings}

\author{Beate Paulus, Krzysztof Rosciszewski$^*$, Peter Fulde}

\address{ Max-Planck-Institut  f\"ur Physik komplexer Systeme,
N\"othnitzer Stra\ss e 38, D-01187 Dresden, Germany}

\address{ ${}^{*}$ Institute of Physics, Jagellonian University, Reymonta 4,
Pl 30-059 Krakow, Poland}

\author{Hermann Stoll}

\address{Institut f\"ur Theoretische Chemie, Universit\"at Stuttgart,
D-70550 Stuttgart, Germany}

\maketitle

\begin{abstract}
We study how the Mott metal-insulator transition (MIT) is affected when we have to deal
with electrons with different angular momentum quantum numbers. For that purpose 
we apply ab-initio quantum-chemical methods to lithium rings in order
to investigate the analogue of a MIT. By changing the interatomic distance 
we analyse the character of the many-body wavefunction and discuss the importance of the $s-p$ orbital quasi-degeneracy within the metallic regime.
The charge gap (ionization potential minus electron affinity) shows a minimum  and the static electric dipole polarizability has a pronounced maximum
at a lattice constant where the character of the wavefunction changes from significant $p$ to
essentially $s$-type.
In addition, we examine rings with bond alternation in order to answer the question 
under which conditions a Peierls distortion occurs.
\end{abstract}

\narrowtext
\section{Introduction}

When the lattice constant of a metallic system is sufficiently enlarged
it will become an insulator, before eventually we end up with a collection of well
separated atoms. This metal to insulator transition was first pointed out by 
Mott([\onlinecite{mott90}] and references therein) and has stimulated  a tremendous amount of work for a better
understanding of it. The standard model for these investigations
has been the Hubbard Hamiltonian\cite{hubbard63}. It assumes one orbital per site
and the electronic Coulomb interactions are reduced to an on-site Coulomb integral
$U$. By applying different approximations Hubbard was able to quantify the conditions
for the occurrence of a M-I phase transition. For that reason it is often called
a Mott-Hubbard transition. More recently, the dynamical mean-field theory
(DMFT) has been applied to the Hubbard Hamiltonian with one electron per site 
and has further elucidated details of the transition 
(for an overview and further references see, e.g., [\onlinecite{gebhard97}]).\\
The purpose of the present work is to go beyond the Hubbard model
in order to understand better how other features might influence the
transition in a realistic system. Lithium is a metal with a strong contribution
of $p$ electrons to the conduction band. With increasing lattice
constant the ratio of $p$ and $s$ electrons is changing
and in the atomic limit we are dealing with a pure $1s^22s$ atomic state. One of the
questions we would like to have answered is the effect of low-energy atomic excitations
from $2s$ to $2p$ on the M-I transition. In order to come close to a 
realistic description of metallic systems we investigate a ring of Li atoms applying quantum-chemical
methods for a treatment of electronic correlations.\\
In terms of a Hubbard model 
the Li-Li distance would correspond to the hopping term $t$. Instead of
only an on-site Coulomb repulsion as in the Hubbard model we take here into account all
possible Coulomb matrix elements. But in the limit of large seperations of atoms 
only the on-site Coulomb repulsion remains.\\ 
The largest ring which we can treat at the multi-reference configuration-interaction (MRCI) level, without excessive computations, 
consists of 10 Li atoms.
A medium-size basis set consisting of Gaussian type orbitals (GTO's) is used in the calculations.
Usually the onset of insulating behaviour is characterized by the appearance of a gap.
However, with a finite number of Li atoms in the ring we always have a finite
energy difference between consecutive energy levels. Therefore, when the atomic distance
is changed we must look for a change in the pattern between the highest occupied molecular
orbital (HOMO) and the lowest unoccupied orbital (LUMO).     
These quantities are only accessible at the mean-field level. At the
correlated level, we select three quantities 
which can give insight as regards the transition behaviour. MRCI methods 
can yield very good approximations to the exact ground-state wavefunction.
The character of this wave function will be the first criterion.
A second quantity which we calculate is the one-particle energy gap defined as the energy difference
which is obtained if an electron is added to, or removed from the system.
As a third quantity, we determine the static dipole polarizability of the system by
applying a small homogeneous electric field. These three quantities are determined 
as functions of the nearest-neighbour Li-Li distance in the ring, and we discuss how the
metal-insulator transition shows up in them.
For these quasi-one-dimensional systems, the question arises whether we can
gain energy by dimerization (Peierls distortion).
We address this question by allowing for a bond alternation.\\ 
In the next section we will discuss some
technical details (Sect. II). In Sect. III, we present the results of different mean-field
methods, for Li rings with increasing number of Li atoms and with increasing Li---Li distance.
The behaviour of the many-body wavefunction is discussed in Sect. IV. Gap energies are
presented in Sect. V and polarizabilities in Sect. VI. 
The Peierls distortion is considered in Sect. VII and conclusions follow in Sect. VIII.

\section{Technical details}

The application of quantum-chemical {\em ab-initio} methods to Li rings is at the focus of the present work.
Starting from a mean-field Hartree-Fock (HF) description, we re-optimize the valence wavefunction in
multi-configuration self-consistent-field (MCSCF) calculations, keeping the $1s$ core electrons frozen at the HF level.
Thereby, the number of valence active-space orbitals is chosen equal to (or larger than) the number of Li atoms of the system,
and all possible occupations of these orbitals are accounted for.
\begin{table}
\begin{tabular}{|l|c|c|c|c|} 
\hline
Basis set &IP (a.u.)&EA (a.u.)&$\alpha$ (a.u.) &$d_{\text dimer}$ ({\AA})\\
\hline
aug-cc-pVDZ&0.196&0.0151&166.0& 2.73\\
aug-cc-pVTZ&0.196&0.0176&167.2& 2.69\\
aug-cc-pVQZ&0.196&0.0209&167.5& 2.68\\
\hline
$[4s]$&0.196&0.0019&0.000 & 3.03\\
$[3s1p]$&0.196&0.0150&164.8 & 2.85\\
$[4s1p]$&0.196&0.0212&165.6 & 2.85\\
$[4s2p]$&0.196&0.0216&165.6 & 2.73\\
$[5s2p]$&0.196&0.0218&166.6 & 2.67\\
\hline
exp.&0.198\cite{IP}&0.0227\cite{IP}&164.8\cite{knight85} & 2.67\cite{IP} \\
\end{tabular}
\caption{ \label{basis}
ACPF values of the ionization potential (IP), the electron affinity (EA), and the dipole polarizibility ($\alpha$)
of the Li atom are listed for different basis sets. 
Corresponding values for the bond length ($d_{\text dimer}$) of the Li$_2$ molecule are listed, too.}
\end{table}
With this choice, we can properly describe the dissociation limit where each atom has one valence electron.
On top of the MCSCF calculations, we apply the multi-reference averaged coupled pair functional (MRACPF) method \cite{gdanitz88,ACPF}.
Here, all configuration-state functions are included which can be generated from the MCSCF reference wavefunction by means
of single and double excitations from the active orbitals.
These calculations are performed with the program package
MOLPRO\cite{molpro2002,wk1,wk2}.\\
The GTO one-particle basis set used for the MCSCF/MRACPF calculations of the present work
is derived from the correlation-consistent polarized valence double-zeta (cc-pVDZ) basis set of Dunning and co-workers\cite{dunning89}.
From this set, we select the three $s$ functions and the first contracted $p$-function.
To describe the negatively charged system, we add an even-tempered diffuse $s$-function (exponential parameter: 0.0107)
resulting in a $[4s1p]$ basis set. We checked the quality of this basis by
calculating the electron affinity (EA), the ionization potential (IP) and the dipole polarizability
of the Li atom (Table \ref{basis}). The $p$ function is essential to describe the
EA and the polarizability properly.
Whereas the IP is weakly dependent on the size of the basis set used, because correlation effects are absent (core-valence
correlation is neglected),
\begin{table}
\begin{tabular}{l|c|c|}
$a=a_0$&EA&IP\\
\hline
HF&0.0208&0.1554\\
MCSCF&-0.0027&0.1672\\
ACPF&0.0335&0.1812\\
FCI&0.0321&0.1820 \\
\hline
$a=1.5 a_0$&EA&IP\\
\hline
ACPF&0.0539&0.1530\\
FCI&0.0500&0.1548 \\
\hline
$a=2.0 a_0$&EA&IP\\
\hline
ACPF&0.0448&0.1703\\
FCI&0.0409&0.1707 \\
\end{tabular}
\caption{\label{meth}
The EA and IP of Li$_6$ are listed for different levels of
approximations and different Li-Li distances.}
\end{table}
the EA substantially increases when a diffuse $s$-function is added. A second diffuse $s$-function or a diffuse
$p$-function have nearly no influence. The polarizability only slightly depends
on the basis-set quality beyond cc-pVDZ, if the basis set contains at least one $p$ function.
The errors of the selected $[4s1p]$ basis set as compared with experimental data
are 1\% for the IP, 7\% for the EA and 0.5\% for the polarizability.
Thus, the $[4s1p]$ basis although being relatively small can describe with sufficient accuracy
all quantities we are interested in and is taken as default basis set in the subsequent calculations. \\
The Li---Li distance in 3-dimensional Li metal is $a_0$=3.03\AA\cite{kittel}, while
the equilibrium distance of the Li$_2$ dimer (2.67\AA\cite{IP}) is smaller by about 10\% . 
For the Li$_{10}$ ring, we calculate an equilibrium distance of 3.06{\AA} which is very similar to the bulk nearest-neighbour
distance. Note, however that the $[4s1p]$ basis overestimates the Li$_2$ bond length by 0.18 {\AA} (the lack of a second $p$ function is
mainly responsible for this error, cf.\ Table \ref{basis}); if this bond-length error carries over to Li$_{10}$, we
would have an equilibrium distance of around 2.9 {\AA} for this system.

\section{Mean-field description}

As a first step of our MIT study for Li rings, we performed mean-field (Hartree-Fock and density-functional (DFT)) calculations 
for various ring lengths and internuclear distances.
With the Crystal program\cite{crystal98} we can do HF calculations for the one-dimensional infinite Li chain.
As basis set we selected the optimized [3s2p] basis set for the three-dimensional metal\cite{doll99}.
\begin{figure}
\begin{center}
\psfig{figure=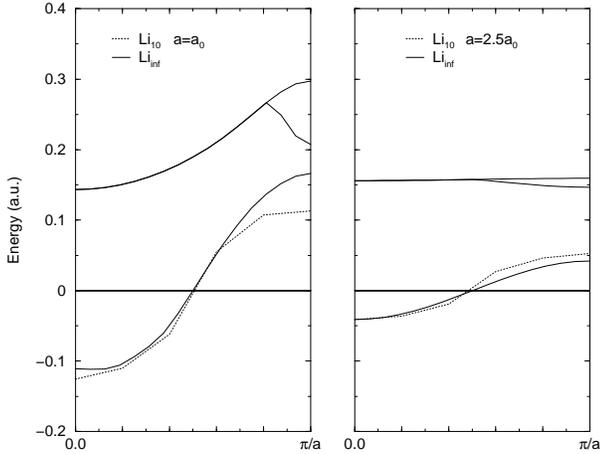,angle=-90,width=8cm}
\end{center}
\caption{\label{bandst} Hartree-Fock band structure for the infinite Li chain for two different 
Li---Li distances. 
The Fermi energy is set to zero.
In addition, a band structure obtained from the levels of Li$_{10}$ is plotted (dashed curve);
for comparison, the center of the $2s$ band is shifted to zero.}
\end{figure}
In Fig.\ref{bandst}, the Hartree-Fock (HF) band structure is shown for the bulk equilibrium internuclear distance $a_0$, and
for $a=2.5a_0$ which corresponds to the dissociation limit.
In both cases the Fermi energy lies within the $2s$ band, but the band width is significantly reduced
with increasing Li---Li distance. The predominant $2p$ band is very flat near the dissociation limit
and well separated from the $2s$ band, whereas for a distance $a_0$ there is no
gap between the $2s$ and $2p$ band.\\
\begin{figure}
\begin{center}
\psfig{figure=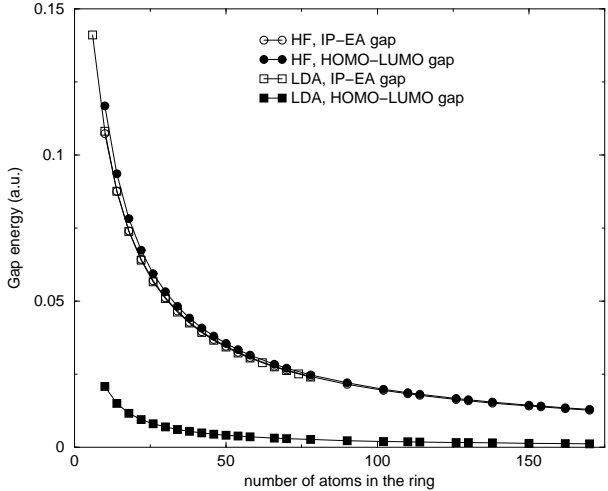,angle=-90,width=8cm}
\end{center}
\caption{\label{hfgap} Mean-field results (HF and LDA) for finite Li rings, in dependence of
the number of atoms in the ring. The Li---Li distance is fixed at the metallic value ($a = a_0$).}
\end{figure}
For finite systems we applied the Turbomole {ab-initio} program package\cite{turbomole} to obtain mean-field results
for very long rings up to $n=170$. We explicitly imposed spatial-symmetry restrictions on our closed-shell
wavefunctions in order to avoid symmetry breaking. (For rings with  $n>$30,  the HF approximation favours a symmetry
broken ground state.) If we generate a HF band structure from the orbital energies of a Li$_{10}$
ring and shift the Fermi energy to zero (dashed curve in Fig.\ref{bandst}), the band width and band shape
correspond well to that of the infinite chain, for both $a=a_0$ and $a=2.5a_0$.\\
The energy gap of Li rings with different
number of atoms, at a fixed Li---Li distance of $a=a_0$,  is shown in Fig. \ref{hfgap}.
In a mean-field approach, the gap energy can either be calculated as HOMO-LUMO gap in the neutral system
or by adding and subtracting an electron to the system, i.e., as difference of the IP and EA.
At the HF level both approaches (approximately) coincide according to Koopmans' theorem\cite{koopmans33}.  
Slight differences occur for smaller rings due to relaxation of
the orbitals when the electron number is changed. This effect is missing in the HOMO-LUMO gap.
Calculating the IP-EA gap  with density functional theory in the local-density approximation
(LDA)\cite{LDA}, the results coincide well with the HF results. Note that in LDA the HOMO-LUMO gap 
is much smaller.
\begin{figure}
\begin{center}
\psfig{figure=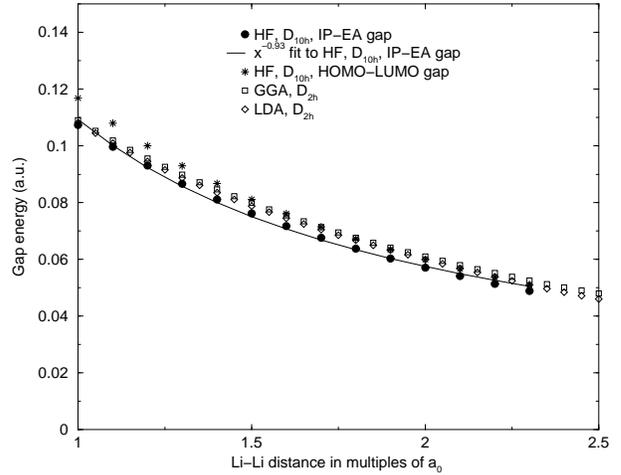,angle=-90,width=8cm}
\end{center}
\caption{\label{li10hfgap} Mean-field results (HF in D$_{\rm 10h}$  symmetry,
DFT(LDA and GGA) in D$_{\rm 2h}$  symmetry) for a Li$_{10}$ ring, in dependence of
the Li---Li distance. The gap energy is taken as the difference of IP and EA; in the HF case, also the HOMO-LUMO gap value is given.}
\end{figure}
This gap is directly related to an 
excitation energy other than the IP-EA gap.
The decay of the IP-EA gap with increasing number of atoms in the ring is slow ($n^{-0.76}$).
(If we would regard the finite ring as consisting of free electrons in a box, the decay should be
proportional to (box-length)$^{-1}$.)
Despite the slow closing of the gap, even for Li$_{10}$ it lies well within the band width of the infinite chain.\\
For the Li$_{10}$ ring, we also calculated the behaviour of the IP-EA gap with increasing
Li---Li distance, from $a_0$ to $2.5a_0$ (Fig.\ref{li10hfgap}). For the mean-field
approaches, HF and DFT (both LDA\cite{LDA} and GGA\cite{GGA1,GGA2}),
the gap is monotonously closing. Thereby, the differences between the DFT variants and the differences between DFT and HF
are not larger than that between the HOMO-LUMO and IP-EA values at the HF level.
(Small differences may also arise due to the use of different formal symmetries (D$_{2h}$ for DFT, D$_{10h}$ for HF).)
If we would model the ring by free electrons in a box again, the decay of the HOMO-LUMO gap should be
proportional to (box-length)$^{-2}$. This is in contrast to an approximate $a^{-1}$ decrease obtained 
in our mean-field calculations, which 
shows that the electrons are far away from the free-electron limit.
It should be kept in mind, however, that the correct dissociation limit cannot be described
with these methods, anyway.
Only multi-configuration methods can properly describe both the dissociation limit and the metallic behaviour. 
Such calculations form the main part
of our present  work and will be discussed below.

\section{The characteristic features  of the ground-state wavefunction}

Taking the example of the Li$_{10}$ ring,
we want to study how the character of the many-body ground-state
wavefunction changes when enlarging the Li---Li distance from
$a=a_0$ to $a=2.5a_0$. For that purpose we performed a MCSCF calculation\cite{MCSCF1,MCSCF2} for the lowest singlet state,
on top of a closed-shell
HF calculation like that described in the previous section. (Using the MOLPRO package, we had to formally change to 
D$_{2h}$ symmetry, however.)
The selection of the active space
is a decisive step: It is well known that for a correct description of 
the dissociation limit in Li$_{10}$  
ten atomic $2s$ orbitals are needed, but for the metallic
regime ($a\approx a_0$) it is not clear how many orbitals are really important.
Therefore we performed a pilot MCSCF calculation for $a = a_0$ with a high number of active orbitals
and selected those natural orbitals whose
occupancies were above a certain 
threshold.
\begin{figure}
\begin{center}
\psfig{figure=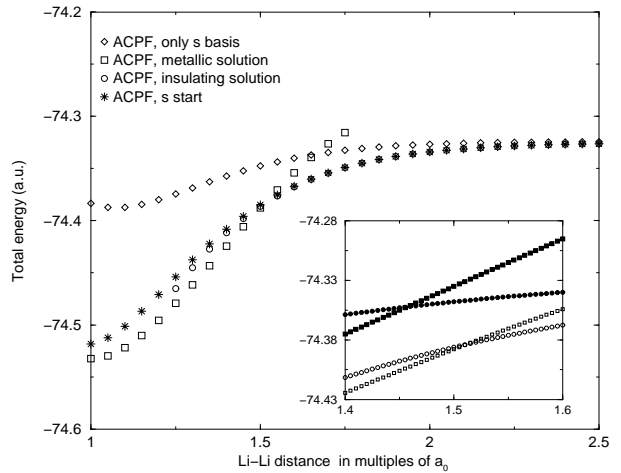,angle=-90,width=8cm}
\end{center}
\caption{\label{wf} MRACPF energies of the Li$_{10}$ ring versus the 
Li---Li distance. The curves labeled  "insulating wavefunction" and "metallic wavefunction" refer to different
MCSCF zeroth-order wavefunctions used as reference for the subsequent MRACPF, cf.\ text.
In the inset an enlarged part of the transition region is shown.
The filled symbols correspond to the MCSCF values. The curves labeled "s start" and "only s basis" refer to  calculations
leaving out p functions in the MCSCF zeroth-order wavefunctions and at both the MCSCF and MPACPF levels, respectively.}
\end{figure}
\begin{figure}
\begin{center}
\psfig{figure=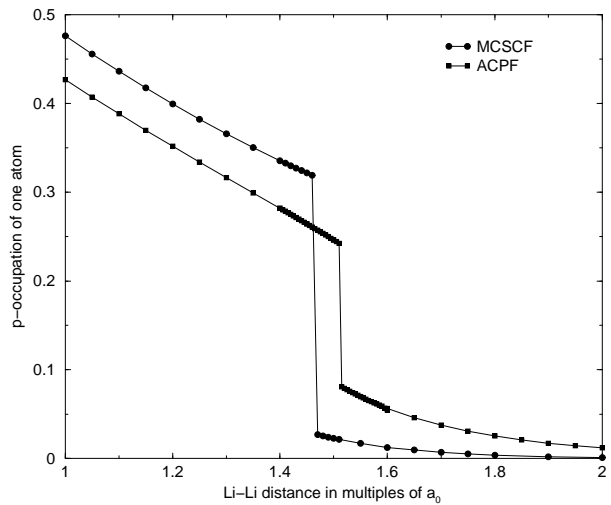,angle=-90,width=8cm}
\end{center}
\caption{\label{pocc} The $p$ occupation of the Li atom in a Li$_{10}$ ring
versus the Li-Li distance. The $p$ occupations are calculated with the Mulliken 
population analysis.} 
\end{figure}
Retaining orbitals with 
occupancies larger than 0.18 and rejecting all the others which have occupancies smaller than 0.03
resulted in 10 active orbitals, too.
However, only
the 9 energetically lowest ones fall into the
same irreducible representations 
(in D$_{2h}$ symmetry) as the 9 lowest-lying orbitals of the insulating regime, while
the highest natural orbital of the metallic and insulating regime differ in symmetry. 
Therefore, we selected for our final MCSCF calculation a 
new set of  11  active orbitals which was generated as the union of the two sets of orbitals described above.
These orbitals were reoptimized at the MCSCF level. Finally, we performed a MRACPF calculation\cite{gdanitz88,ACPF} on top
of the MCSCF, in order to include dynamical correlation effects as fully as possible.
Note that the multi-reference treatment is important in 
both limits: in the metallic regime to describe static correlation with the
quasi-degenerate $2p$ orbitals, and in the 
insulating regime to get the dissociation into degenerate $2s$ orbitals right.\\
Although we included in the active space all orbitals which are important in 
the limits of small and large atomic distances,
the total energy
as function of the Li---Li distance is not a smooth curve (Fig. \ref{wf}).
This is so because the nature of the active orbitals changes along the curve.
When starting the calculations from large $a$ (wavefunction for the insulating state) and always using
the previous solution as starting point for the next smaller lattice constant, this yields
a slightly different solution in the region of the MIT than when starting from the metallic regime and
increasing $a$.
The difference is larger, of course, for the MCSCF energies  than at the MRACPF level;  
With a full CI (i.e., including all higher excitations) the differences would vanish.
However, a 
single smooth curve can be obtained even at the MRACPF level 
if it is based on a MCSCF calculation with only $s$ functions in the active space. 
In that case $p$ excitations appear at the MRACPF level only. It is clear
that such a procedure favours the insulating solution:
it yields a ground state energy for $a=a_0$ which is by $\sim$0.02 a.u.
higher than the value obtained with variationally optimized MCSCF orbitals.
Still, the so obtained ground-state potential curve is qualitatively correct also in the metallic regime.
It is interesting to compare it with a MRACPF curve obtained by leaving out p functions in the one-particle basis set altogether.
Here, we still observe a (shallow) minimum near $a=a_0$, but the energy is higher by more than 0.1 a.u.
Thus, $p$ contributions amount to about two thirds of the binding energy of the Li$_{10}$ ring.
\\
Turning back now to our most accurate potential curve:
The change of the character of the MRACPF wavefunction as well as that of the underlying MCSCF wave function as a 
function of $a$ is seen best by
a Mulliken population analysis for the $p$ orbitals (Fig. \ref{pocc}). For the "metallic" MCSCF solution
the atomic $p$ orbital occupancy is nearly 0.5 at $a=a_0$ and slowly drops to 0.3 at $a \approx 1.5 a_0$, while 
for the "insulating" solution
it is less than 0.03 for $a \geq 1.5 a_0$. Even in the MRACPF case, the $p$ population changes
by a factor of $\sim$3. Taking this rather dramatic change as a signature of the MIT 
we find for all investigated rings  a region around $1.5 a_0$ , where the character of the
wavefunction changes significantly.

\section{The one-particle energy gap}

\begin{figure}
\begin{center}
\psfig{figure=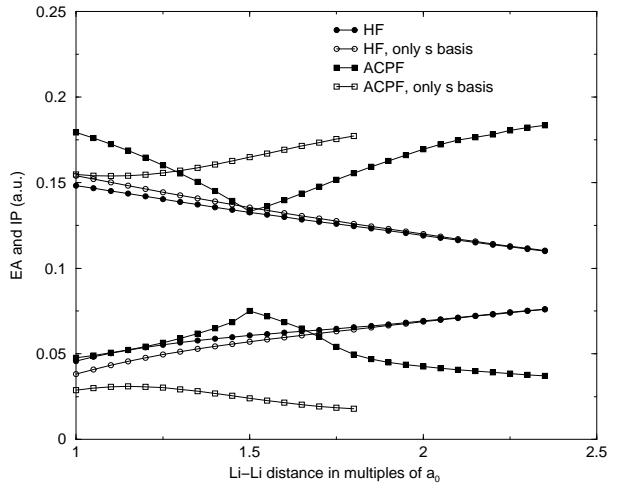,angle=-90,width=8cm}
\end{center}
\caption{\label{li10gap} HF and MRACPF values for the EA and IP
of a Li$_{10}$ ring,
versus the Li-Li distance. For comparison, we add results from calculations without $p$ functions in the basis set ("only s basis").
}
\end{figure}

The energy gap of an insulator is determined by the energy differences between the ground state of
the neutral system and the ones with one electron added and subtracted.
In the case of a Li ring, we
calculate the MRACPF ground-state energies of the Li$_n^-$, Li$_n$ and Li$_n^+$ systems
and determine  EA=E$({\rm Li}_n)$-E(${\rm Li}_n^-)$ and  IP=E(Li$_n^+$)-E(Li$_n$) and
from IP-EA the corresponding gap. 
For a metallic solid adding and removing an electron 
cost an energy given by the chemical potential. 
In finite systems, there always remains the difference between the one-particle energies
of the additional and the missing electron as well as the influence of relaxation and correlation effects.\\
The EA and IP for Li$_{10}$ are plotted in Fig. \ref{li10gap}.
In the metallic regime, 
the HOMO-LUMO gap is expected to decrease with increasing Li-Li distance 
exactly like it happens 
for free electrons in a box with changing length. 
\begin{figure}
\begin{center}
\psfig{figure=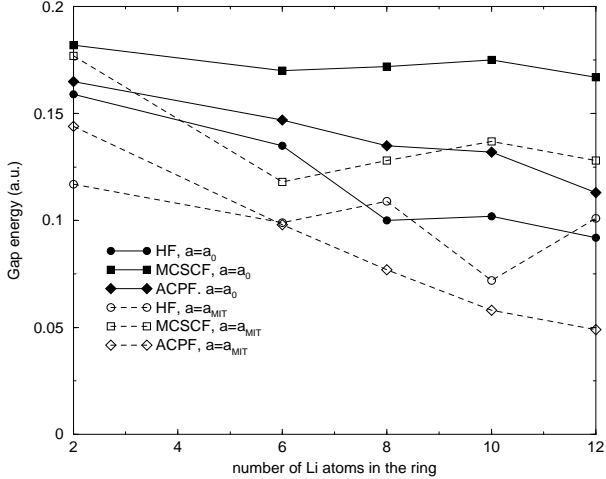,angle=-90,width=8cm}
\end{center}
\caption{\label{scaling} HF, MCSCF and MRACPF one particle energy gaps are plotted versus the number of Li atoms in the ring
for two different lattice constants, cf.\ text.}
\end{figure}
In fact, at the HF level we obtain a monotonous shrinking of the gap, both with our standard basis set ([4s1p])
and with a pure s basis set ([4s]).
The availability of p polarization functions stabilizes the charged systems as compared to the
neutral one, leading to a slightly smaller gap for $a=a_0$ than with a pure s basis, but the difference
goes to zero for $a \rightarrow \infty$ and the general trend of a closing gap with enlarging Li-Li nearest-neighbour
distance is the same in both cases.
However, the trend completely changes when correlation effects are included. The most important change
concerns the $a \rightarrow \infty$ limit, where HF fails to describe correctly the dissociation.
Instead of a closing of the gap, the correct limit is the difference between the atomic IP and EA.
This limit is obtained indeed, in our MRACPF calculations. As already noted in Sect. 2, the atomic EA is severely
underestimated in the case of a pure s basis set, but this is efficiently remedied with our
standard [4s1p] basis set. 
We furthermore note that the EA converges much
slower to the atomic limit than IP (EA($a=2.35a_0$)=1.6 EA(atom),IP($a=2.35a_0$)=0.94 IP(atom)).
We think that this is due to the larger size of the Li$^{-}$ ion than of the Li$^+$ ion.
Interestingly, the variation of IP and EA from $a=a_0$ to $a \rightarrow \infty$ is not completely monotonous.
Already for the case of a pure s basis, the gap is nearly constant (and even slightly decreases) in the
range $a_0 < a < 1.3 a_0$, and only then gradually opens up to the atomic limit.
Including p contributions leads to a significant increase of IP and EA for $a=a_0$ as well as to a distinctly
non-monotonous behaviour of the gap as a function of $a$, with a minimum around $a=1.5 a_0$.
The effect is due to a stabilization of the charged systems, with respect to the neutral one, by means
of polarization clouds around holes/electrons.
Thus, for a finite ring in the metallic regime the
gap is closing when enlarging the lattice constant, 
at the MIT point it is minimal,  
then it starts to increase gradually   approaching 
the atomic limit value.  \\
We investigated the gap behaviour not only for the Li$_{10}$ ring, but also 
for the Li$_2$ dimer, for the Li$_6$ and Li$_8$ rings, and for the Li$_8$ cube. (Li$_4$ arranged in a square has a 
negative EA; hence, the calculation of the one-particle energy gap does not make any sense. Li rings with odd number of
Li atoms do not have a closed shell HF ground state and therefore are not considered here.) For Li$_{12}$,
we  performed calculations at $a=a_0$ and $a=a_{\rm MIT}^*$, where $a_{\rm MIT}^*$
is defined as the point where the character of the MCSCF wavefunction changes from significant $p$ character
to an essentially $s$ one.
For all investigated ring systems, the minimal value of the gap 
occurs for nearly the
same lattice constant in different rings ( between 1.42$a_0$ and 1.55$a_0$).
This distance coincides approximately with the one where  
the character of the wavefunction changes.
In Fig.\ref{scaling}
the energy gap at $a=a_0$ and $a=a_{\rm MIT}$ is listed for different methods 
and for different ring lengths.
When increasing the number of atoms in the ring, 
the minimal gap energy at the ACPF level decreases faster than the gap at $a=a_0$. 
Unfortunately, the 
behaviour of the gap for much longer rings cannot be calculated by the methods we use  because it is
computationally too expensive. 
The gap calculated at the MCSCF level seems to saturate for longer rings, i.e. only the 
dynamical correlations are closing the gap. 
As yet, we cannot make a reasonable extrapolation and find out
for which ring length a zero gap is to be expected. But for all investigated rings
we found similar behaviour with increasing Li-Li distance independent of the number of Li atoms in the ring.
In spite of this we reemphasize that
although we have a finite system instead of a real metallic system, 
we still see a transition from a metallic-like regime
(where the gap is closing with increasing Li-Li distance) 
to an insulating regime where the gap is 
opening towards the atomic limit.\\
In order to validate the methods used in the present paper, we performed
full-CI (FCI) calculations\cite{knowles84} for Li$_6$ at three different lattice constants, i.e.,
we determined the lowest variational energies in each case which can be obtained with our standard [4s1p]
one-particle basis set. The resulting EA and IP are listed in
Table \ref{meth}. The ACPF values coincide well with the FCI results and
therefore are expected to provide a reliable, quantitative description of the gap energy.\\
Finally, let us note, that
we observe a similar minimal gap for the Li$_8$ cube, 
where no sharp jump in the $p$ occupation pattern of the wave
function is visible. In the cube,
the minimal gap occurs at a somewhat larger distance (1.65$a_0$) than it does in the ring and 
is also a little bit larger (0.09 a.u.), but the qualitative 
behaviour remains the same.

\section{Dipole Polarizability}
\begin{figure}
\begin{center}
\psfig{figure=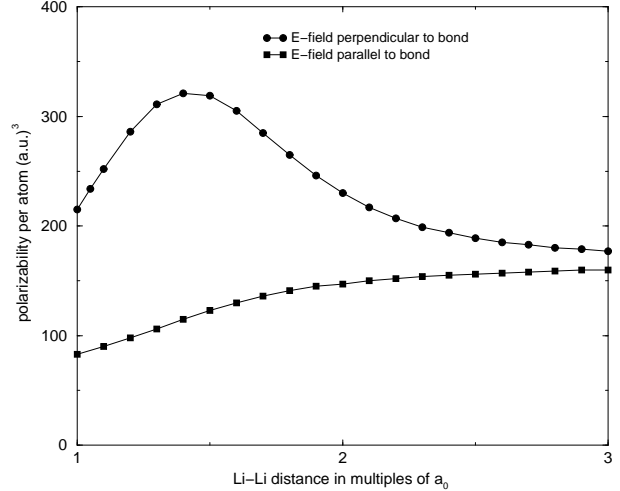,angle=-90,width=8cm}
\end{center}
\caption{\label{li2pol} The MRACPF static dipole polarizability of the Li$_2$ dimer, for the electric field
parallel and perpendicular to the bond direction, is plotted versus the Li---Li distance.}
\end{figure}
As a third quantity which can be used to define the MIT, we calculate the static electric dipole polarizability of the
system. We apply a static electric field of strength up to 0.035 a.u. and
perform a quadratic fit for the energy of the system, $E({\mathcal{E}})=E(0)-\frac{1}{2}\alpha{\mathcal{E}}^2$, yielding the
polarizability $\alpha$. Higher order terms can be neglected for the small field strengths we chose.
For a metallic system the polarizability should be infinite, therefore a steep increase of the
polarizability can indicate an insulator-metal transition.\\
As a test example, we selected the Li$_2$ dimer
and applied the $\mathcal{E}$ field both in the bond direction and
in a direction perpendicular to that, while gradually enlarging the Li---Li 
distance (Fig.\ref{li2pol}).
\begin{figure}
\begin{center}
\psfig{figure=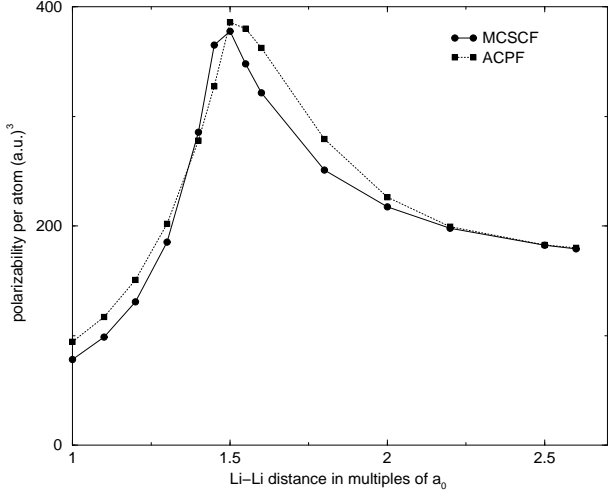,angle=-90,width=8cm}
\end{center}
\caption{\label{li10pol} The static dipole polarizability of the Li$_{10}$ ring, for the electric field
in the ring plane, is plotted versus the Li---Li distance for two different methods.}
\end{figure}
At the equilibrium bond length, we obtain an average polarizability of 118 a.u. (the experimental value is 111  a.u. 
\cite{antoine99}).
If the electric is parallel to the bond 
direction, the polarizability exhibits  a maximum
at a Li---Li distance similar to that where the minimal gap occurs. 
For a field direction perpendicular to the bond,
the polarizability increases monotonously to two times the atomic 
value as the HOMO-LUMO gap becomes smaller.\\
For the Li$_{10}$ ring, we find the same behaviour as in the dimer 
when we apply the field perpendicular to the
ring plane. When we apply the field 
in the ring plane we find a pronounced maximum.
In Fig. \ref{li10pol}, we plot the polarizability 
per atom for the Li$_{10}$ ring, both  at the
MCSCF and MRACPF levels. 
A crossing of the curves occurs within the transition region, since we have  two different solutions
at the transition.
 MCSCF and MRACPF polarizability values differ only slightly.
At $a=a_0$ the calculated MRACPF value is 94.4 a.u.. 
Experimental data are available for the static dipole
polarizability of a Li$_{10}$ cluster\cite{benichou99}; 
unfortunately the cluster geometry is not fully known, but it can be supposed
to be two- or three-dimensional.
The experimental value is 70.2 a.u. and it 
roughly agrees with our value. 
As already noted,
the polarizability increases with decreasing 
lattice constant (i.e., when coming from the insulating side).  For an infinite system
the polarizability should diverge at the MIT. We see a precursor of this
divergence and can therefore  use it to define a MIT 
even for a finite system. 
The rapid decay of the polarizability in 
both  the metallic  and the
insulating regime is due to the
opening of the gap. 

\section{Lattice distortion}

\begin{figure}
\begin{minipage}[t]{8cm}
\psfig{figure=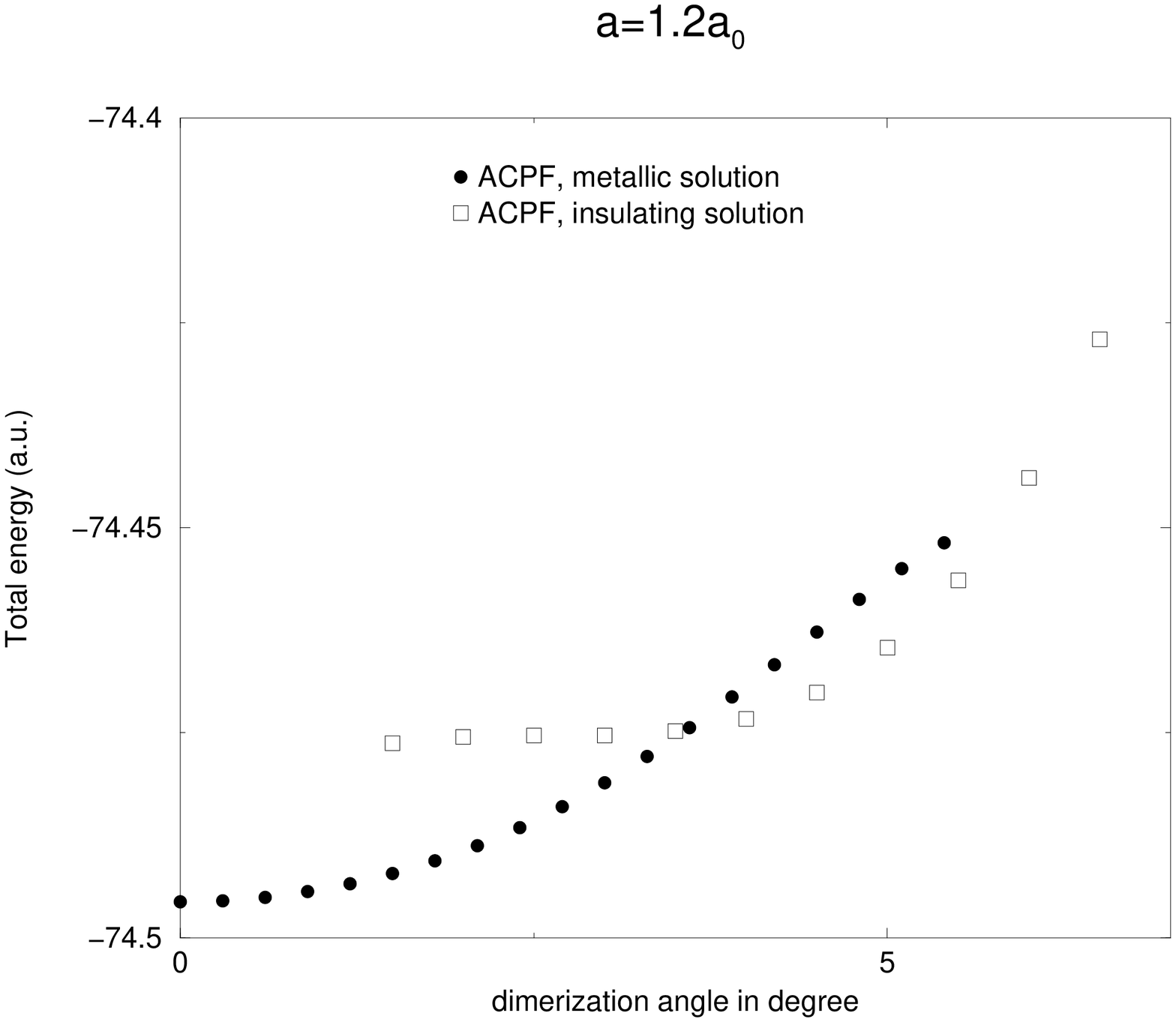,angle=-90,width=6.5cm}
\end{minipage}
\begin{minipage}[t]{8cm}
\psfig{figure=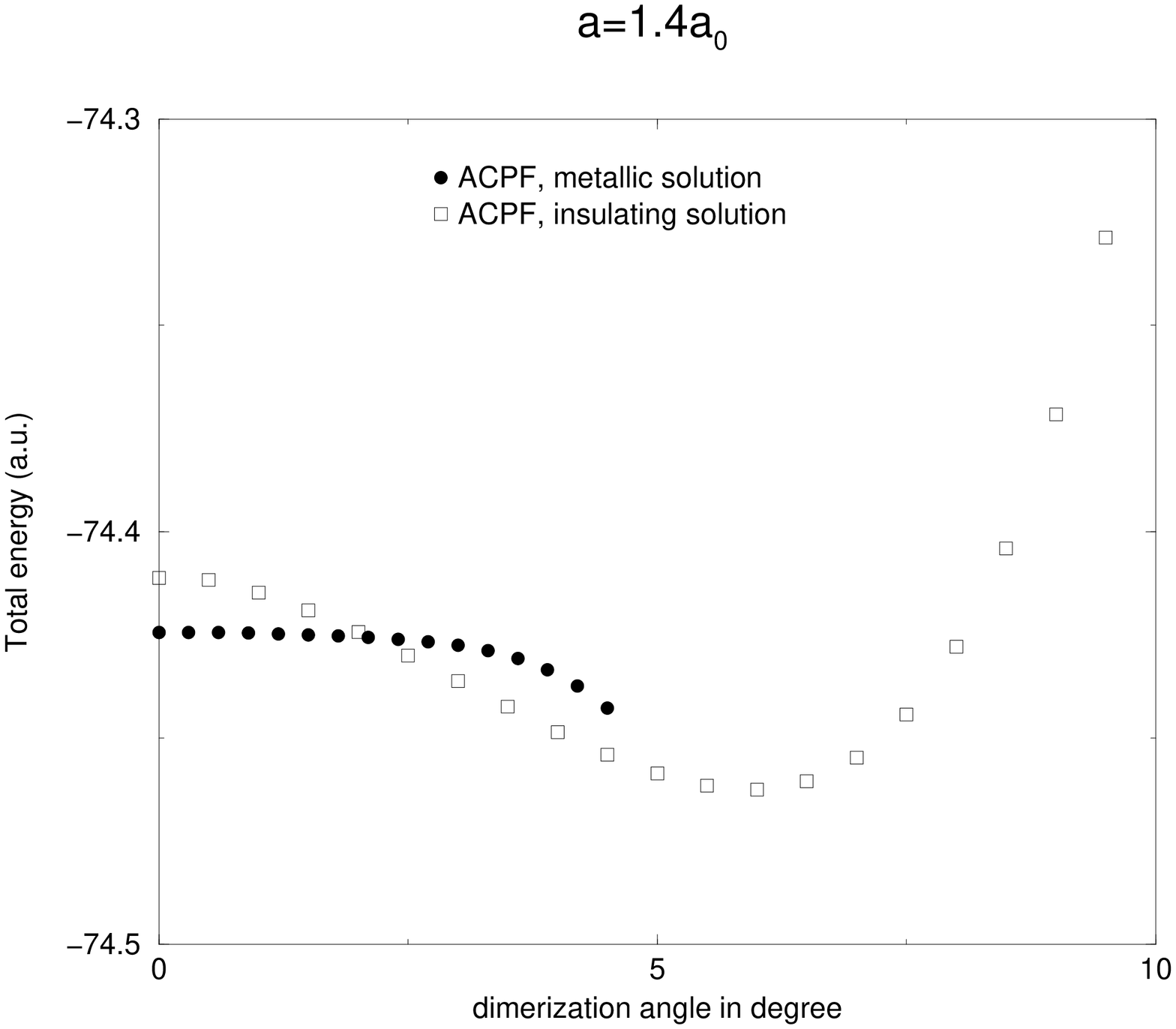,angle=-90,width=6.5cm}
\end{minipage}
\caption{\label{dimer} The total energy of the Li$_{10}$ ring is plotted versus the dimerization angle
(defined as deviation from the ideal 36$^{\rm o}$ angle in Li$_{10}$, in degree)
for different average lattice constants.}
\end{figure}

In this last section, we allow for a bond alternation along the ring, so that the Li atoms can form dimers.
For average lattice constants larger than that where the M-I transition occurs (i.e., for lattice constants where the insulating
solution is the ground state) the formation of dimers is energetically favoured. The resulting dimer internuclear distance
agrees well with the calculated distance of the free dimer (2.88 \AA). The energy gain is due 
to bond formation between two Li atoms. In the metallic regime ($a\le a_{\rm MIT}$) 
a Peierls distortion\cite{peierl55} occurs, stabilizing
the dimerized state (see Fig \ref{dimer}). 
However, decreasing the lattice constant further, we found a point ($a=1.3 a_0$ for Li$_{10}$),
below which the equidistant arrangement is 
the ground state, which would imply a metallic state for the infinite chain. \\
At the HF level, a bond alternation is found for all (average) lattice constants;  
electronic correlations seem to suppress this bond alternation at the equilibrium lattice constant.
This is  in contrast to
poly-acetylene, where correlations reduce the bond alternation without fully suppressing it
\cite{koenig90}. 
 
\section{Conclusions}

We investigated the analogue of the metal-insulator transition
for one-dimensional lithium, applying high-level quantum-chemical ab-initio methods. 
The aim of this investigation was to find out how the MIT is modified from the one in the Hubbard 
model, when we have to deal with several orbitals per site. We have chosen Li as an example since
here the different orbitals correspond to different angular momentum quantum numbers, i.e., $s$ and
$p$ electrons. By treating a Li$_{10}$ ring we could apply accurate quantum-chemical ab initio methods.
Following the original ideas of Mott,
a metallic system will become an insulator when the lattice constant is sufficiently large.
But in distinction to the Hubbard model it is here the interplay
of the $s$ and $p$ orbitals which strongly affects the MIT.
At the transition point ($a_{\rm MIT}\approx 1.5 a_0$) the character of the many-body wavefunction changes from significant $p$
to essentially $s$ character. 
Therefore it must be kept in mind that in a real solid the re-population of orbitals 
belonging to different angular momenta may have similarly strong influence on the MIT as 
changes in the ratio of the Hubbard $U$ to the kinetic energy.
We found that approximately at the same interatomic distance where the $p$ character
of the wavefunction changes so strongly, the one-particle
energy gap is minimal and the static electric polarizability has a maximum.
Increasing the ring length, the minimal gap closes faster than the gap at the equilibrium distance;
the reduced level spacing for both longer rings and larger lattice constants enhances dynamical correlation effects.
Allowing  for a bond alternation in the ring, the system dimerizes for all (average) lattice constants
larger than $\sim 1.3 a_0$), for the insulating case due to Li$_2$ molecule formation,
for the metallic case due to  Peierls distortion. For even smaller lattice constants, the correlations
stabilize the equidistant arrangement of the Li atoms along the ring. \\
Further investigations are planned 
for open boundary conditions
(i.e., linear fragments of the infinite chain) and 
for other alkali metals (thereby, e.g. mixing different
alkali atoms to mimic the ionic Hubbard model).
  
\begin{appendix}
\section{List of abbreviations}
\begin{tabular}{ll}
MIT& metal-insulator transition\\
HOMO& highest occupied molecular orbital\\
LUMO& lowest unoccupied molecular orbital\\
GTO& Gaussian type orbital\\
cc-pVDZ& correlation consistent polarized valence double-zeta\\
HF& Hartree-Fock\\
MCSCF& multi-configuration self-consistent field approximation\\
MRCI& multi-reference configuration interaction\\
MRACPF& multi-reference averaged coupled pair functional\\
IP& ionization potential\\
EA& electron affinity\\
DFT& density functional theory\\
LDA& local density approximation\\
GGA& generalized gradient corrected approximation\\
FCI &full CI -- configuration interaction with all excitations
\end{tabular} 
\end{appendix}

\end{document}